\documentclass[pra,aps,showpacs,twocolumn,groupedaddress,superscriptaddress,tightenlines]{revtex4}
\usepackage{graphicx}
\usepackage{amsmath}
\usepackage{mathrsfs}
\usepackage{textcmds}
\allowdisplaybreaks[1]
\newcommand{\be}{\begin{equation}}
\newcommand{\ee}{\end{equation}}
\newcommand{\bea}{\begin{eqnarray}}
\newcommand{\eea}{\end{eqnarray}}
\newcommand{\ket}[1]{\left|#1\right\rangle}
\newcommand{\bra}[1]{\left\langle #1\right|}

\newcommand{\bc}{\begin{center}}
\newcommand{\ec}{\end{center}}

\renewcommand{\(}{\left(}
\renewcommand{\)}{\right)}
\renewcommand{\[}{\left[}
\renewcommand{\]}{\right]}
\newcommand{\forget}[1]{}

\newcommand{\re}{{\rm e}}
\newcommand{\ri}{{\rm i}}
\newcommand{\rd}{{\rm d}}
\begin{document}
\title{Extreme subwavelength atom localization  via coherent population trapping}
\author{Girish S. Agarwal\footnote{On leave:
Physical Research Laboratory, Navrangpura, Ahmedabad-380 009,
India}}
\affiliation{Department of Physics, Oklahoma State University, Stillwater,
OK-74078}
\author{Kishore T. Kapale}
\affiliation{Quantum Computing Technologies Group, Jet Propulsion Laboratory,
California Institute of Technology, Mail Stop 126-347, 4800 Oak Grove Drive, 
Pasadena, California 91109-8099}
\date{\today}
\begin{abstract}
 We demonstrate an atom localization scheme based on
monitoring of the atomic coherences.  We consider atomic transitions in a Lambda configuration where the control field is a standing-wave field. The probe field and the control field produce coherence between the two ground states. We show that this coherence has the same fringe pattern as produced by a Fabry-Perot interferometer and thus measurement of the atomic coherence would localize the atom.  Interestingly enough the role of the cavity finesse is played by the ratio of the intensities of
the pump and probe.  This is in fact the reason for obtaining extreme subwavelength localization. We suggest several methods to monitor the  produced localization.
\end{abstract}
\maketitle

\section{Introduction}
Precision position measurement of an atom has been of interest since the early days of quantum mechanics. An illustrative example is the Heisenberg microscope which allows measurement of the position of an atom by observing  scattering of light from it. Modern tools of quantum optics have made such thought experiments a  reality.  A variation of the Heisenberg's microscope has been studied in Ref.~\cite{Stokes:1991}; it allows suboptical wavelength position measurements of moving atoms as they pass through the optical fields. Since then a variety of methods have been studied for subwavelength localization of an atom passing through standing-wave fields. Subwavelength atom localization has been shown to be possible through quantum interference~\cite{Paspalakis:2001}, by measuring a quadrature phase of the light field in a cavity as the atom passes through it~\cite{Storey:1992}, through detection of spontaneously emitted photons~\cite{Kien:1997, Holland:1996}, by diffracting the atoms through a measurement induced grating~\cite{Kunze:1997} or by Raman-induced resonance imaging~\cite{Gardner:1993}. Recently phase control of subwavelength atom localization has also been shown giving rise to variety of interesting controllable features in atom localization~\cite{Sahrai:2005}.  Similar techniques could also be used to perform measurement of the atomic center of mass wavefunction as proposed recently by one of us and collaborators~\cite{Kapale:2003}. 

The study of atom localization affords practical applications in the area of nanolithography at the Heisenberg limit~\cite{Johnson:1998,Boto:2000} along with its fundamental importance to the areas of atom optics~\cite{Adams:1994}, and  laser cooling and trapping of neutral atoms~\cite{Phillips:1998}. In this article we propose a high resolution  localization scheme based on the phenomena of coherent population trapping (CPT)~\cite{Arimondo:1996}. A standing-wave control field and a weak probe field cause the atom to evolve into a long-lived superposition of the two ground states. Measuring the position dependent coherence of this trapping state localizes the atom.
Extending the scheme to two dimensions, optical lattices with tighter than usual confinement at each lattice point can be obtained. Such a strongly confined lattice structures could be useful to study several predictions of the Bloch theory of solids and Mott transitions in much cleaner systems.

The article is organized as follows. First we discuss the scheme of localization through the phenomena of coherent population trapping. As the localization of the atom occurs through the detection of coherence we provide a couple of detection schemes that would lead to localization of the atom in a subwavelength regime as it passes through the standing wave field. To ascertain the localization we  present the momentum distribution of the localized atom and point out unique features that, if seen experimentally, would confirm the localization of the atom in the subwavelength domain. Next, we point out several applications of the scheme and the unique advantages it offers. Finally, we present our conclusions.

\section{Localization Scheme}
\begin{figure}[ht]
\includegraphics[width=0.9\columnwidth]{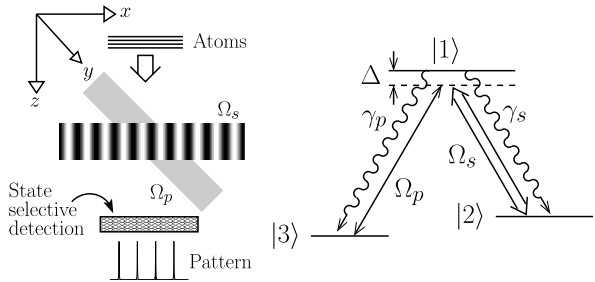}
\caption{\label{Fig:LevelScheme} Application of a strong standing-wave field, on $\ket{1}-\ket{2}$ transition, and a weak Probe field, on $\ket{1}-\ket{3}$ transition, prepares the atom in a particular position-dependent superposition of the states $\ket{2}$ and $\ket{3}$ in steady state. Probing this superposition through selective measurement of the population of state $\ket{2}$, causes subwavelength localization of the atom at the nodes of the standing-wave field.}
\end{figure}

The level scheme for our model is depicted in Fig.~\ref{Fig:LevelScheme} (cf.~Ref.~\cite{Kunze:1997}) along with a possible experimental geometry.  The atom has $\Lambda$ type level configuration, such that it is coupled to two fields in two-photon resonance Raman configuration as shown in the figure. A possible atomic specie could be $^{85}$Rb with the levels $\ket{2}$ and $\ket{3}$ drawn from the two hyperfine components of the ground state. The strong control field is the standing-wave field that could be a field confined in an optical cavity or a Fabry-Perot resonator or produced using two counter-propagating laser beams.  The Rabi frequencies of the standing-wave field ($\Omega_s(x)=\Omega_s \sin(kx)$, with the wavenumber $k = 2 \pi/\lambda$) and the weak probe field ($\Omega_p$) could be of the order of tens of MHz and the laser power level could be few mW/cm$^2$ as in a typical CPT experiment~\cite{Adams:1994}. The laser fields are along the $x-y$ plane as shown in the Fig.~\ref{Fig:LevelScheme}, such that the atom sees them at the same time at is travels along $z$ direction.

The initial state of the atoms is  $\ket{3}$ as they enter the fields in the transverse $z$ direction with the center-of-mass distribution uniformly spread over a few standing-wave field wavelenghts.  We assume that the velocity of the atoms in the $z$ direction can be treated classically and there is no significant variation of the $x$-velocity of the atoms as they interact with the fields; therefore the kinetic part of the atoms can be ignored within the Raman-Nath approximation (RNA)~\cite{Meystre:1999}.  This poses limitation on the extent of the localization possible as stricter confinement along $x$-direction causes increasing spread in the $x$-momentum of the atom. However, RNA requires the recoil kinetic energy,  $\hbar^2 k_{a}^2/2 m$, acquired by the atom of mass $m$ due to its $x$-momentum, $k_{a}$, should be significantly less than the interaction energy with the fields, $\hbar \Omega_p$ and $\hbar \Omega_s$, i.e.,
\begin{equation}
\frac{\hbar^2 k_{a}^2}{2 m} \ll \hbar \Omega_p, \hbar \Omega_s\,,
\end{equation}
throughout the interaction region. 

 The interaction of the fields with the three-level atom, within the two-photon resonance condition, can be described through the interaction Hamiltonian
\begin{equation}
\mathscr{H} =  - \hbar (\Omega_p \ket{3}\bra{1}  + \Omega_s(x) \ket{2}\bra{1} ) \re^{- \ri\, \Delta t} + \text{H.~c.}\,.
\end{equation}
\forget{The corresponding density matrix equations can be written as
\begin{equation}
\dot{\rho} = \frac{\ri}{\hbar} [\mathscr{H}, \rho] - \sum_{i = 2, 3}\frac{\gamma_{1i}}{2} (\ket{1}\bra{1} \rho - 2 \rho_{11}\ket{i}\ket{i} + \rho \ket{1}\bra{1})\,,
\end{equation}
with $\gamma_{12}=\gamma_{s}$ and $\gamma_{13} = \gamma_{p}$. In principle, the  solutions of the density matrix equations can be obtained at steady state.}
A quick observation shows that the state
\begin{equation}
\ket{\Psi}=(\Omega_p \ket{2} - \Omega_s(x) \ket{3})/\Omega\,,
\end{equation}
where $\Omega= \sqrt{|\Omega_p|^2 + |\Omega_s(x)|^2}$,
does not evolve dynamically as $\mathscr{H}\ket{\Psi} = 0$. Thus, an atom initially prepared in state $\ket{3}$ will end up in state $\ket{\Psi}$  at steady state, if the two-photon resonance is maintained. As the population in the state $\ket{\Psi}$ can not escape, it is termed as the trapping state and the phenomena is called coherent population trapping. This state, being a superposition of ground states, does not decohere. There could be a physical loss of atoms during the transit time; however, this is not a limitation as these atoms would not be localized. \forget{Considering flat-top intensity distributions of the laser beams, steady state could be easily reached for the atom during its flight-time through the lasers.} With the detuning $\Delta$ sufficiently large, the atom is hardly excited and recoil due spontaneous emission becomes a non-issue~\cite{linewidth}. 

Thus, ensuring the two-photon resonance, introduces coherence $\rho_{23} = -\Omega_p \Omega_s^*(x)/\Omega^2$ between the levels $\ket{2}$ and $\ket{3}$ that carries the  spatial dependence of the standing-wave field. Also, as the CPT state $\ket{\Psi}$ is reached, the population of the state $\ket{2}$ is given by 
\begin{equation}
\label{Eq:rho22}
\rho_{22}(x) = 1/(1 + \mathscr{R}\sin^2 kx),
\end{equation}
where $\mathscr{R}= |\Omega_s|^2/|\Omega_p|^2$. Therefore, so long as the steady state is reached within the interaction time (typically 100 ns) governed by the $z$-velocity of the moving atoms, monitoring the population of state $\ket{2}$ is a sufficient measure of the coherence of the CPT state. This interaction time is well below the limit of $<7 \mu$s required to remain within the RNA for $^{85}$Rb atoms undergoing optical transitions~\cite{Kunze:1997}. Our numerical study, considering both Gaussian and flat-top beams, shows that the spatial intensity profile is inconsequential if the trapping-state is reached in steady-state within the interaction time. 

It is important to point out that the  atom passing through the standing-wave field and the probe field will see spatial variation of the laser intensity. This can, equivalently, be seen as the time variation of the interaction Rabi frequencies as experienced by the atom. To study  the dependence of the laser intensity variation at the beam boundaries as seen by the atom we studied two types of line-shapes, namely Gaussian and flat-top:
\begin{align}
\Omega(t) &= \Omega_0 \exp\[\(\frac{t-t_0}{w}\)^2\] \quad \text{Gaussian}\,, \nonumber \\
\Omega(t) &= \frac{\Omega_0}{2}\[\tanh\(\frac{t-t_1}{r_1}\) -\tanh\(\frac{t-t_2}{r_2}\)\]\quad \text{flat-top}\,.
\end{align}
Here, $t_0$ and $w$ denote the center and the width of the Gaussian respectively and  $t_1$, $r_1$, $t_2$ and $r_2$ correspond to the start of the rise of the beam intensity, rise time, start of the intensity fall, and the fall time respectively for the flat-top beam. For both  the Gaussian and the flat-top beams the maximum intensity is denoted by  $\Omega_0$.
Our numerical studies, choosing either kind of beam shape for both the standing-wave and the probe fields with the maximum intensity ratio chosen to be $\mathscr{R}$ show that as long as the beam parameters introduced above are chosen  such that the atom reaches the steady state while well before the beam intensity starts to fall, the CPT state is reached in the steady state. Once the steady state is reached, as there are no strong decoherence mechanisms, beam intensity fall does not change the steady state. 

Physically, the above mechanism can  be thought of as state preparation. 
During state preparation, time varying light-fields could be applied to reach a desired state of the atoms. Once the desired state is reached the light fields are turned off. Nevertheless, turning off the fields after the state is created does not change the state until it decoheres through some other mechanisms. In the above, if the beam parameters are appropriately chosen so that the steady state can be reached while the atom is in the beams, then CPT state is reached and turning off of the beams does not change the state.

\begin{figure}[ht]
\centerline{\includegraphics[scale=0.39]{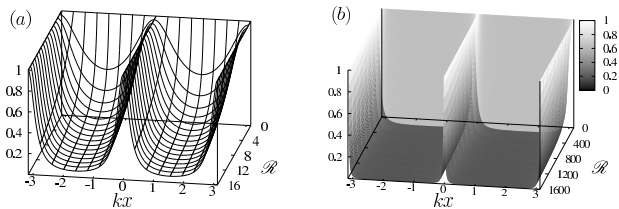}}
\caption{\label{Fig:Plots} Emergence of subwavelength localization as the ratio $\mathscr{R}$ is increased. Plot of $\rho_{22}(x)$ vs $kx$ for various values of $\mathscr{R}$. The range of $kx$, $\{-\pi,\pi\}$, covers one  wavelength of the 
standing-wave field. $(a)$ For $\mathscr{R}=0$, $\rho_{22}$ is uniform at all spatial points $k x$. $(b)$ Increasing the cavity field strength localization peaks emerge with decreasing peak width.}
\end{figure}

We plot the population $\rho_{22}$ versus $kx$ in Fig.~\ref{Fig:Plots} for various values of the ratio $\mathscr{R}$. 
The population shows peaks at the nodes of the standing-wave field. The FWHM of the peaks is given by $k\, \Delta x = 2/\sqrt{\mathscr{R}}$. Thus, for small values of the ratio of the Rabi frequencies the peaks are not well defined. In fact, for $\mathscr{R}=0$,  $\rho_{22}$ has no spatial dependence. The small $\mathscr{R}$ behavior is shown in the part $(a)$ of the figure where the emergence of the localization peaks is clearly seen. For large $\mathscr{R}$, the peaks become quite sharp, leading to subwavelength localization.

It can be noted that Eq.~\eqref{Eq:rho22} has the same structure as the transmission function of the Fabry-Perot cavity~\cite{Born:1999}, with the ratio $\mathscr{R}$ of the effective field intensities playing the role of the cavity finesse. Interestingly, in the present model the finesse can be controlled by varying the relative intensities of the control and probe fields, leading to much sharper features in $\rho_{22}$ (see Fig.~\ref{Fig:Plots}).

It is imperative to point out the meaning of the result obtained above. The state vector of the atom, including its center-of-mass degree of freedom and the CPT internal state, can be written as
\begin{equation}
\ket{\Psi_{\rm COM}} = \int  f(x) \ket{x} \frac{1}{\Omega}\(\Omega_p \ket{2} - \Omega_s(x)\ket{3}\) \,    \rd x \,,
\end{equation}
where $f(x)$ is the center-of-mass wavefunction. Thus, observance of a peak in $\rho_{22}$ amounts to a collapse of the internal state of the atom to $\ket{2}$ at one of the positions corresponding to the nodes of the standing-wave field. This leads to the collapse of the center-of-mass wavefunction at one of those positions leading to localization of the atom.  With sharper features in $\rho_{22}(x)$ one obtains deeper localization of the atom within a subwavelength region of the optical field. Our treatment assumes that $f(x)$ is uniform over the standing-wave field, which remains unchanged even after the interaction with the optical fields. It is the measurement process that leads to the localization of the atoms. It is important  that the atoms are detected in a state selective manner after their interaction with the laser fields. If, however, the total population of the states $\ket{2}$ and $\ket{3}$ is measured, there would be no position information in the result obtained. This illustrates the role of quantum coherence and interference in the CPT state. 

\section{Detection leading to localization}
In the following we discuss methods for detection of the CPT state. As alluded earlier simplest possible and sufficient technique is to measure $\rho_{22}$. Alternatively, measuring the coherence $\rho_{23}$, which is proportional to $\sin{kx}$, can also be used to detect the CPT state. Thus, any measurement technique monitoring the coherence $\rho_{23}$, would observe sharp decrease in the coherence if the atom passes through the nodes of the standing-wave field, leading to localization of the atom. At all other positions the coherence $\rho_{23}$ carries a non-zero value. However, ascertaining a zero of a function experimentally could be challenging as opposed to observing peaks.  Therefore, we discuss an alternative technique to measure  the coherence without having to ascertain a zero value.

\begin{figure}[ht]
\centerline{\includegraphics[scale=0.33]{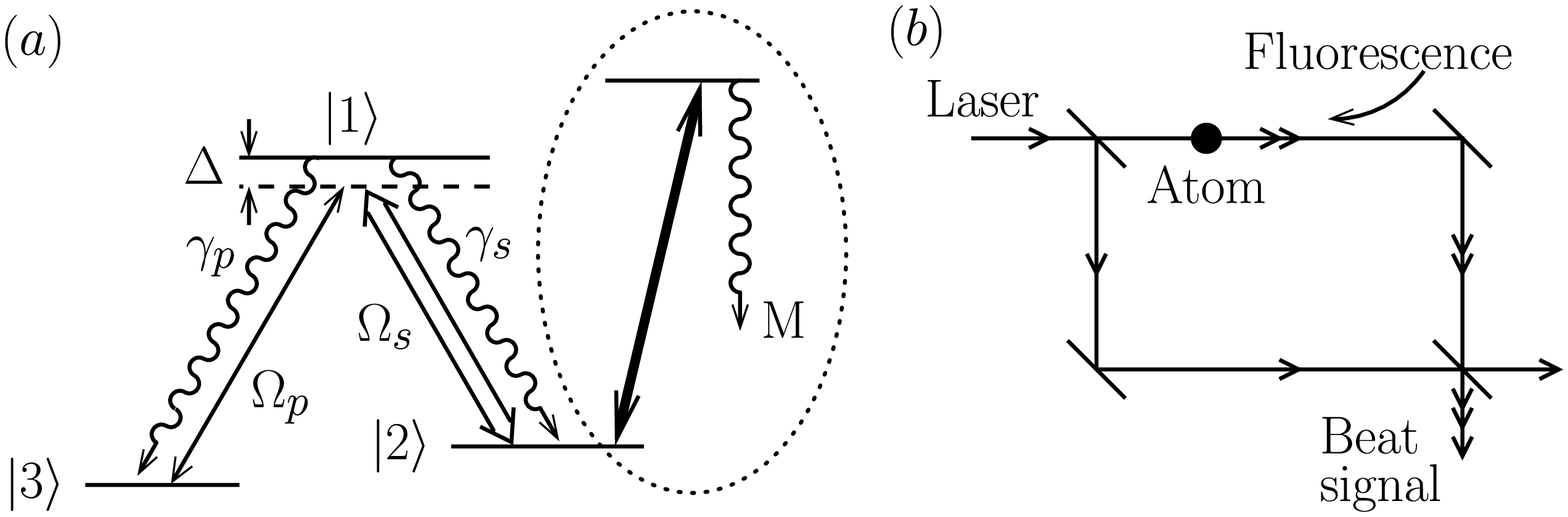}}
\caption{\label{Fig:Monitor} Techniques to monitor the atom localization. The  CPT state of the atom could be probed by monitoring the population of the state $\ket{2}$, i.e., $\rho_{22}(x)$. $(a)$  Fluoresence shelving technique for efficient measurement of $\rho_{22}(x)$. Applying a strong shelving field and monitoring the resulting fluoresence  leads to localization of the atom at the nodes of the standing-wave field.  $(b)$ Heterodyne measurement of the fluoresence from the atom also leads to localization as soon as the heterodyne beat signal is detected.}
\end{figure}

\forget{In the following we discuss methods for detection of the CPT state through the measurement of either the population $\rho_{22}$ or the coherence $\rho_{23}$.}
Monitoring the population, $\rho_{22}$, of state $\ket{2}$ can be accomplished by various techniques, we describe two interesting ones. The first method is based on the fluoresence shelving techniques described in Ref.~\cite{Blatt:1988}. It involves applying a strong drive field (See Fig.~\ref{Fig:Monitor}$(a)$) to selectively couple the to-be-measured state to a short lived excited state. Then monitoring the fluoresence  from the excited states gives 
information about the population of the state to be measured. 
Alternatively, another high-resolution technique, based on the heterodyne measurement of fluoresence from a single atom developed by the Walther group~\cite{Hoffges:1997}, can be devised. The scheme is depicted in Fig.~\ref{Fig:Monitor}$(b)$. A weak probe acts on the transition involving state $\ket{2}$ and any other excited state of the atom besides $\ket{1}$. Heterodyning the resulting fluorescence with the incoming laser gives information about the population of the state $\ket{2}$.

Monitoring the coherence, $\rho_{23}$, can be accomplished by illuminating the atom, as it comes out in a CPT state, by a $\pi/2$ radio-frequancy (rf) field directly coupling the states $\ket{2}$ and $\ket{3}$. Such a pulse gives a transformation $\ket{3}\rightarrow  (\ket{2}+\ket{3})/\sqrt{2}$ and $\ket{2}\rightarrow  (\ket{2}-\ket{3})/\sqrt{2}$. Then, monitoring the population of state $\ket{2}$, which can be shown to be $P_2=(1/2) - \Omega_p\, \Omega_s \sin{(kx)}/\Omega^2$,  using the techniques depicted in Fig.~\ref{Fig:Monitor} gives the measure of coherence prepared by the control and the probe fields.  $P_2$ has sharp features at the nodes of the standing-wave field for larger values of $\mathscr{R}$ and has a value $P_2=0.5$; measuring it leads to atom localization.  Fluorescence based measurement techniques are useful in optical domain as the recoil caused by the emission of a photon is negligible.

\section{Momentum Distribution of the localized atoms}

\begin{figure}[ht]
\includegraphics[scale=0.32]{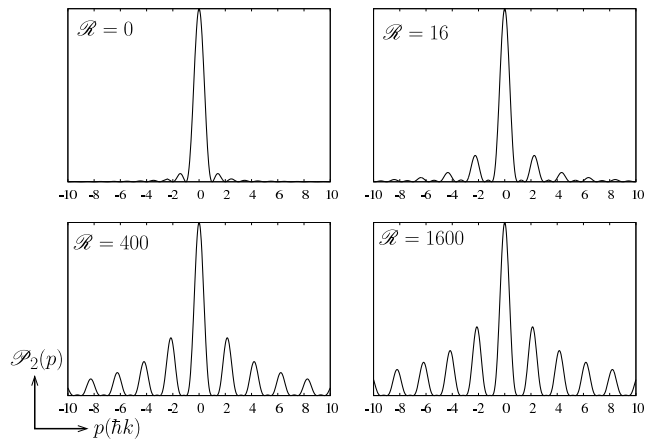}
\caption{\label{Fig:MD-Plots} Momentum distribution $\mathscr{P}_2(p)$ in the far-zone, for atoms in state $\ket{2}$, for different values of parameter $\mathscr{R}$.}
\end{figure}

In a real experiment, as in Ref.~\cite{Kunze:1997}, the momentum distribution of atoms would be measured. To predict the experimental results expected from our model we plot the momentum distribution,  along the $x$ axis, obtained by evaluating 
\begin{equation}
\mathscr{P}_2(p)=|\langle p, 2 | \Psi_{\rm COM}\rangle|^2 = \Bigl|\int_{-\pi}^{\pi} \rho_{22}(x)\,\re^{\ri p x} \rd x\Bigr|^2
\end{equation} in Fig.~\ref{Fig:MD-Plots}. 
Here we have assumed  that $f(x)=1$, i.e. uniform distribution of the atoms along the standing-wave field. Moreover, the integration is confined to a single wavelength of the standing wave. These assumptions do not cause loss of generality  and the conclusions arrived at are 
The plots show the momentum distribution of the atoms after they are detected in state $\ket{2}$ after interacting with the probe and a single wavelength of the standing-wave field. For sharper localization,  resulting due to increased $\mathscr{R}$, the higher momentum components start emerging in accordance with the Heisenberg uncertainty relation. The number of peaks for the plot corresponding to no localization are large in number compared to other plots, despite the amplitude being negligible for high $p$. As  the localization peaks start appearing, alternate peaks in the momentum distribution start diminishing (See for example plots for $\mathscr{R}=0$ and $\mathscr{R}=16$). Thus, observing this feature in the momentum distribution is a clear signature of the localization of the atom. As  the finesse parameter is increased higher momentum components start becoming prominent. Therefore, to remain within the Raman-Nath approximation, the highest momentum order $p\approx10$ observed for $\mathscr{R}=1600$ should satisfy $p^2 \hbar k_{a}^2/2m\ll \Omega_p$. For optical transitions of alkali atoms, $\hbar k_{a}^2 / 2 m $ is in the kHz range whereas $\Omega_p$ is in the MHz range (Table I in Ref.~\cite{Adams:1994}), giving $p=10$ to be well-within the Raman-Nath approximation.

\begin{figure}[ht]
\includegraphics[scale=0.4]{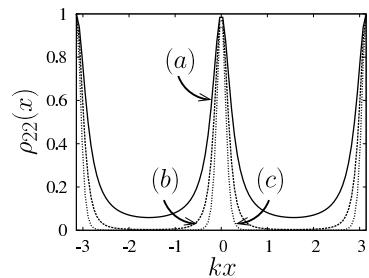}
\caption{\label{Fig:Repeats} Sharper localization by repetition of the localization zones with a moderate finesse parameter $\mathscr{R}=16$. $(a)$, $(b)$ and $(c)$ correspond to one,  two and four localization zones arranged one after the other.}
\end{figure}

As seen above, stricter localization of the atoms gives rise to larger momentum spread. Thus, it is sometimes advisable to have smaller finesse parameter but to repeat the localization procedure a number of times. We observe that sharper localization is also possible by increasing the localization zones, instead of increasing the value of the finesse parameter. To illustrate, we choose a moderate value, $\mathscr{R}=16$, for the finesse parameter and show the localization peaks for single, two and four localization zones arranged one after the other in Fig.~\ref{Fig:Repeats}.  It is clearly seen that as the number of zones are increased, sharper position localization is obtained. This could be a more practical way of achieving sharper localization than increasing the strength of the standing-wave field relative to the probe.

\section{Applications and unique advantages}
We suggest some  applications and advantages  of our scheme: It can be noted that ac-Stark shift induced potential on a two-level atom due to two detuned, and counter-propagating light beams gives rise to optical lattice for trapping neutral atoms~\cite{Jessen:1996}. The cold neutral atoms in their ground states are trapped at the nodes of the standing-wave fields arising due to the counter-propagating beams. This optical-lattice  trapping potential, which is proportional to $\sin^2 (kx)$ where 
$k$ is the wavenumber,  is much broader than the localization feature attainable through Eq.~\eqref{Eq:rho22}. Thus, by applying our localization scheme in two-dimensions~\cite{extension}, deeper and much narrower lattice structures can be obtained. Such experiments could be easily done with degenerate Bose or Fermi gases to study several predictions of the Bloch theory of solids and observation of Mott transitions~\cite{Greiner:2002}. It should however has to be borne in mind that state selective detection of atoms from the ensemble of atoms in a BEC would be required. Whenever, this kind of detection becomes available experimentally our scheme can be readily employed to obtain new kind of optical lattices and atomic Mott insulators.

Noting the spread of the momentum distribution with stricter localization of the atom, any practical implementation of this scheme would require working in the spatial domain such that the atoms are not significantly diffracted.  For example, for lithographic applications, the state selective detection and deposition of the atoms on the substrate would have to be done very close to the interaction region. This may make it necessary to use fibers to collect the fluorescence signal. This, nevertheless, does not pose any serious  limitation  as other state selective deposition techniques could be employed instead.  The standing-wave field can be thought of as a mask for the atoms as opposed to a physical entity used in current lithographic setups. One can also envision two-dimensional~\cite{extension} spatial dependence of the light field giving rise to arbitrary two-dimensional localization pattern of atoms for lithographic applications.

Furthermore, the feature size given by our localization scheme (Eq.~\eqref{Eq:rho22}) can always be chosen to be much smaller that the width of the center-of-mass position distribution of the entering atoms, even if the latter is considerably smaller than the wavelength of the standing light field. This provides a strong advantage for atom localization because,  through monitoring of atomic coherences, we can control atomic distributions which are much smaller than the wavelength of the light fields and produce strong localization.

\section{Conclusions}

To summarize, we have devised a scheme for extreme localization of an atom as it is passing through a standing-wave field, based on the phenomena of coherent population trapping. The resolution of localization peaks can be arbitrarily increased by changing the relative intensity of the probe and standing-wave control fields. The atomic coherence obtained this way resembles the transfer function of Fabry-Perot cavity with controllable finesse.  Increasing the finesse leads to increasing resolution for atom localization. We have discussed several methods to probe the resulting CPT state of the internal atomic states, either through selective monitoring of the population of one of the states or through the coherence measurement. We have also discussed the signature of the localization observable in the momentum distribution of the atoms in the far zone. It is also clarified how the resulting momentum distribution also helps validate the Raman-Nath approximation. We have suggested various techniques for implementation of the model for fundamental as well as practical applications.
\acknowledgments

Part of this work, done by KTK, was carried out 
at the Jet Propulsion Laboratory under 
a contract with the National Aeronautics and Space Administration (NASA). 
KTK acknowledges support from the National Research Council and
NASA, Codes Y and  S. GSA thanks E. Arimondo and G. Rempe for useful discussions.


\begin{thebibliography}{10}
\bibitem{linewidth}
The laser linewidths must be smaller than the spontaneous decay width of the excited state otherwise frequency-locking of the control and probe lasers will become necessary.
\bibitem{extension}
Extension to higher dimensions could be achieved using a pair of counter-propagating laser beams for each spatial axis as in the atom trapping schemes.
\end{thebibliography}

\end{document}